\documentclass[11pt,fleqn]{article}

\usepackage[a4paper,text={6.125in,10.0in},centering]{geometry}
\usepackage{latexsym}
\usepackage{amsmath}
\usepackage{amssymb}
\usepackage{graphicx}
\usepackage{mathrsfs}
\usepackage{bm}
\usepackage{amsthm}
\usepackage[usenames]{color}
\usepackage{subfigure}
\usepackage{cite}
\usepackage{epstopdf} 
\usepackage{authblk}
\usepackage{comment}
\usepackage[official]{eurosym}

\graphicspath{{eps/}{pdf/}}
\graphicspath{ {./Figures/} }

\newcommand{\Pen}{\mathcal{P}}

\begin{document}
\title{A Dynamical Model of the Industrial Economy of the Humber Region.}
\author[1,2]{Christopher J.K.~Knight}
\author[1,3]{Alexandra S.~Penn}
\author[2]{Rebecca B.~Hoyle}

\affil[1]{\small ERIE, Department of Sociology, University of Surrey, Guildford, GU2 7XH, UK}
\affil[2]{\small Department of Mathematics, University of Surrey, Guildford, GU2 7XH, UK}
\affil[3]{\small Center for Environmental Strategy, University of Surrey, Guildford, GU2 7XH, UK}

\maketitle

\begin{abstract}
The Humber region in the UK is a large and diverse industrial area centred around oil refining, chemical industries and energy production.  However there is currently a desire to see the region transition towards a more bio-based economy.  New bio-related industries are being situated in the region as a consequence of policy and economic  incentives.  Many of these industries are connected through their supply chains, either directly, or by sharing common suppliers or customers and the growth or decline of one industry can hence  have impacts on many others.  Therefore an important question to consider is what effect this movement  towards  bio-based industry will actually have on the regional economy as a whole.
In this paper we develop a general abstract dynamical model for the metabolic interactions of firms or industries.  This dynamical model has been applied to the Humber region in order to gain a deeper understanding of how the region may develop.  The model suggests that the transition to a bio-based economy will occur with oil refining losing its dominance to bioethanol production and biological chemical production, whilst anaerobic digestion grows as a major source of electricity, in turn driving up the value of regional waste aggregators and arable farming in the overall economy. 
\end{abstract}

\section{Introduction}
The Humber region is a large diverse industrial area in the UK.  It is centred around the ports of Grimsby, Immingham and Hull on the tidal estuary of the UK's largest river system.  This port complex is one of the largest and busiest in Europe. 
In fact the area surrounding the estuary contains 27\% of the UK's oil refining capacity and  infrastructure for 20\% of national gas landing ~\cite{PKL13}.  The wider region has many diverse industries from farming, to energy production, to heavy industries.   Many of these industries interact with one another, often via the material supply of goods or services.  This can range from the unexceptional use of electricity by almost all industries (in most instances procured from energy suppliers via the grid), to industry-specific needs such as biomass for co-firing power plants.  These interactions form a complicated web representing the industrial economy of the Humber region.  As well as these physical, metabolic interactions there are also what we might call \emph{social} interactions, both positive and negative.  For instance, joint bidding for funding, competition for scarce, highly skilled workers, etc.  We shall not dwell on such social interactions between firms in this paper, however we extend our methods to take into account their effect in~\cite{KPH13}.  Instead we here seek to understand the metabolic interactions and how they influence the region as a whole.

Obviously such an intricate network of relationships is not unique to the Humber region.  In fact, any economy which has a  regional component could  be represented by a complicated set of interconnections between constituent industries.  However not all industrial economies are as complicated as one another.  For instance, consider local economies which are based around one major firm (a hub and spoke district~\cite{M96}).  Such economies are likely to have a relatively simple set of relationships between the industries present; either supplying the major firm or buying and using its products.   Nor are intricate networks of relationships restricted to `man made' economies.  Consider, for instance natural ecosystems, and the complexity of various food webs made up of consumer-resource interactions~\cite{E27,HCI87,BMS05}.

Unlike industrial districts, much research has been carried out on food webs and how to model them in a quantitative, whole system, manner.  In contrast, modelling of industries tends to focus on the supply chains of individual firms~\cite{CH02,DMP09,HBW03} rather than modelling of all firms within an economic or geographic district.  This means, for instance, that competition effects for resources are not completely included.  That is, competition may be included if two firms supply a firm in the supply chain, but not if a competing firm does not supply (in some way) the firm that the supply chain is focussed on.   This could have huge effects, for instance if a competing firm (not in the supply chain) went out of business, this might mean that a firm in the supply chain doesn't have enough business and so also fails, having a knock-on impact on the whole supply chain.  Many of these models of supply chains are static, although some are dynamic, often using multi-agent approaches to simulate supply decisions~\cite{BBP01,SSS98}.  Due to the lack of whole system (high level) dynamical models for interdependencies of firms in an industrial district we draw inspiration from the field of ecological modelling.

The archetype of whole system ecology models is the Lotka-Volterra model for predator prey interactions, originally developed by Lotka in the 1920s,~\cite{L25}.  
This is a dynamical population-based model with the change in population of each individual species being represented by a differential equation.  Each equation contains terms relating to population growth, for instance birth rates, which could be related to the availability of food, and hence to the population of any prey species.  Each equation also contains a term representing the decrease in population due to death.  This term may be dependent on the size of predator populations.  The way that the growth and decay terms depend on the population sizes of other species couples all of the equations together, creating a simple model of the complete food web.  Variants of this basic Lotka-Volterra model have been applied to model many different food webs~\cite{MBN03,T03}, however, to our knowledge, no work has attempted to model the web of industrial firms in the same way.  In this paper we seek to address this deficit by creating a Lotka-Volterra type model for the interaction of industries.  
Instead of each equation modelling the size of a population of a particular species we instead create an equation for the `size' of each firm or industry.  Where `size' can be thought of as an abstract concept which in some way represents the health and wealth of a firm.  The coupling of these equations is then via the `size' of supplier and customer firms rather than prey and predator populations.  This dynamical model of the interactions of industries in a economic or geographic district will be quite high level and general without going into many of the specifics considered when modelling individual supply chains.  The idea behind the model is not to give intricate detail about the system but rather represent it with a `broad brush' in order to make statements on the scale of the whole system, and how it might respond to changes. 
  
In Section~\ref{S:HumberRegion} we apply this model to the industrial economy of the Humber region to learn more about the implications of the interconnected structure of the metabolic connections between industry types.  In particular we investigate how the region may transition from being centred around fossil fuels to become a bio-based economy.

\section{A Dynamical Model of Metabolic Interactions}
An industrial economy is made up of several different firms or industry types interacting in various ways, in a spatially confined region.  As an initial attempt to model the dynamics of such a district of industrial types we shall only consider metabolic interactions, consisting of the trade in goods and services.  Further, we shall make the model as simple as possible so that modifications to include social interactions~\cite{KPH13} may be made and the model remain amenable to analysis.  This means, among other things that the model shall concern only one good (locally), or equivalently make the assumption that all goods (sold within the district) are interchangeable.  Whilst this assumption is questionable in most instances, it hugely simplifies the modelling process and, more importantly, means that it is possible to find the data needed to initialise the model.  It would be possible to make the model more complicated, however this would be at the expense of the model's usability.  It must be remarked that the aim of the model is to express the general behaviour of the system as a whole, as such, the level of assumptions made is appropriate.

\begin{figure}[h!]
\centering
\includegraphics[height=2in]{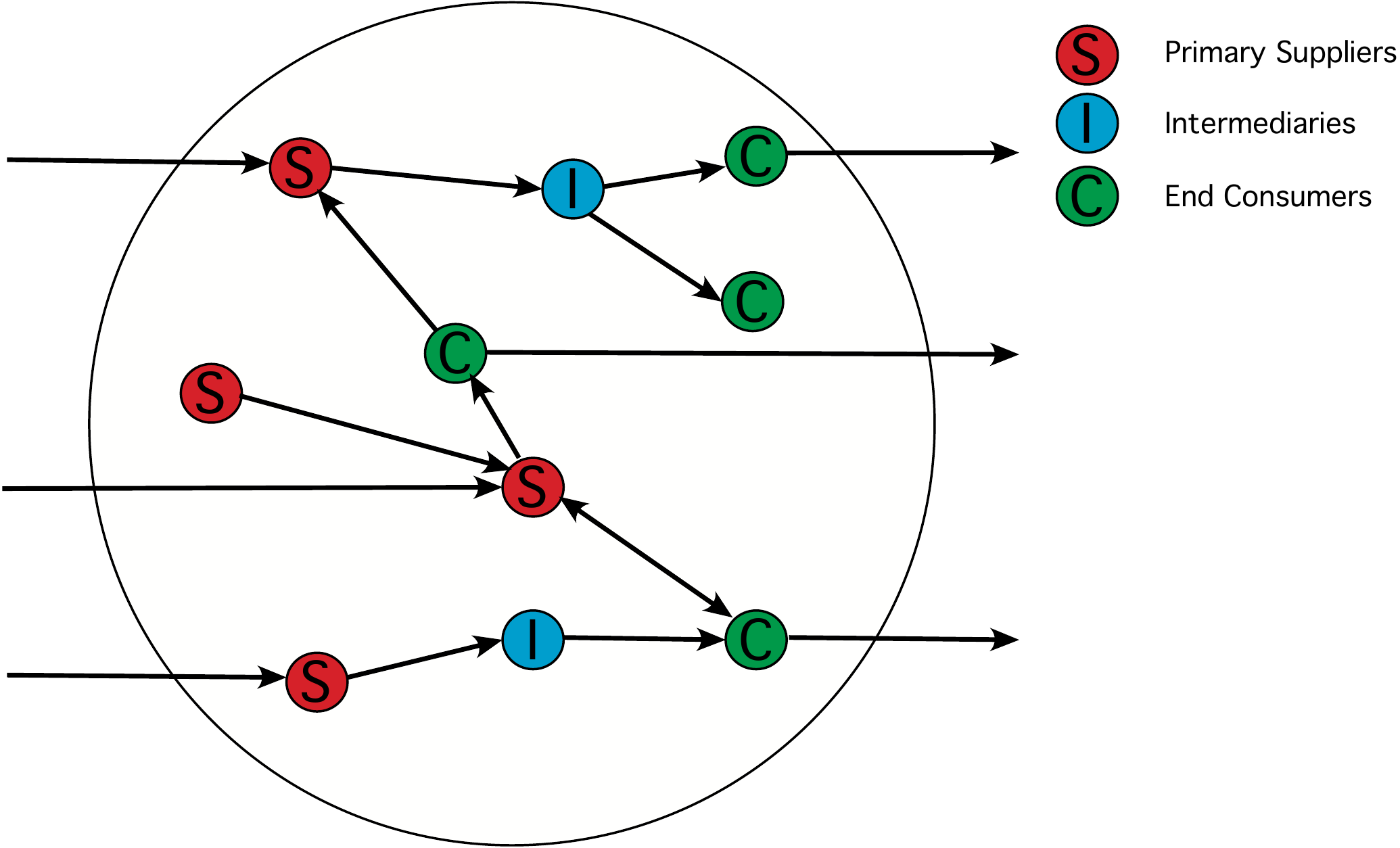}
\caption{An example network with nodes marked as S: `primary Supplier', I: 'Intermediary', C: `end Consumer'. }\label{NTypes}
\end{figure}

As no spatially confined industrial economy (an industrial district) is entirely closed there will always be flows of materials and services into and out of the district.  Similarly unless we model on the scale of the individual there will be transactions between entities in the model and those which have not been modelled, for instance individual consumers.  To account for this edge of the district we split the industry types into three categories signifying their relationship with what lies beyond the district.  They are denoted `\emph{Primary Suppliers}' ({\bf S}), `\emph{Intermediaries}' ({\bf I}) and `\emph{End Consumers}' ({\bf C}).  The distinction relates to the structural role of each firm, both the `primary suppliers' and the `end consumers' have some form of trade link with entities beyond those modelled in the local district.  The rest of the firms are denoted as `intermediaries'.  The `primary suppliers' are those firms who derive product from somewhere outside of the network.  For instance, if a local network is considered then any firm which buys material from further afield is denoted as a `primary supplier'.  Alternatively a firm which mines for resources (as long as the resource itself is not modelled) is a `primary supplier'.
`End consumers' are those firms who sell product beyond the network, this can either be thought as global trade when considering a local network or as selling of product to individual customers (people or small firms not included in the model).      It is important to recognise that a `primary supplier' may also buy from other firms in the network.  Similarly `end consumers' may supply other nodes in the network.   It is also possible for a firm (or industry) to be both a `primary supplier' and an `end consumer', in this case the firm is called a \emph{hub} of the network.  Figure~\ref{NTypes} shows an example of a local metabolic network with labels denoting the type of each node.  Note that every node must be connected (backward) to at least one supplier and connected (forward) to at least one consumer (a node with no outgoing connections to other nodes in the network is automatically  an `end consumer', and one with no incoming connections from other nodes in the network is automatically a `primary supplier').

With this distinction we can start formulating our model for the interactions of industries in a region.  We shall start by considering the industries denoted `intermediary', the equations for industries of the other two classes can be easily derived from them.  Each intermediary (by definition) will have suppliers and customers amongst the other firms (industries) that make up the model.  For a specific firm denoted `firm $i$' the set of firms which supply products or services to it is denoted $S_i$, and the set of firms which buy products or services from firm $i$ is denoted $C_i$.  At its most basic our Lotka-Volterra type model says that each year a firm or industry grows by the amount it sells its products and services for, less the amount it spent buying the raw materials or services it needed, less overhead costs.  However actually writing this basic model down proves to be fairly complicated.

We denote the wealth/health/general utility of firm $i$ by $u_i$, and the worth of the products or services sold to firm $j$ by $G(u_i,u_j)$.  The worth of a product (or service) may be different to those selling it and those buying it.  The firm which sells the product is unlikely to sell it at cost price, meaning that their idea of value must take into account the profit that they made in producing the product or offering the service.  As our model only deals (locally) with one product any feedback loops may cause issues, leading to an exponential rise in the cost of goods as they circle round such a loop.  To avoid this we assume that the value of the goods (as perceived by the buyer) is their actual material and production cost, whilst the value of the goods to the seller is the material and production cost of the product plus the profit they make on it.  This has the effect of renormalising everything at each time step and is similar to inflation adjusting the costs for each firm.  Thus the cost to a firm $i$'s utility of buying from firm $j$ is $G(u_j, u_i)$ and the gain to firm $i$'s utility of selling to firm $j$ is $(1+\epsilon_i)G(u_i,u_j)$.  Where $\epsilon_i$ is the percentage profit that firm $i$ makes.  Finally, if we denote the percentage overhead costs of firm $i$ by $d_i$ then the basis of our model can be written:
\begin{equation}
\dot{u}_i=(1+\epsilon_i) \sum_{j\in C_i}G(u_i,u_j)-\sum_{j\in S_i} G(u_j,u_i) - d_i u_i.
\end{equation}

This equation forms the basis of our model, and is almost complete for intermediary firms.  The one major component still to add is the effect on the utility if a firm is not able to buy all of the materials or services it requires.  This could affect the amount of its own product a firm is able to sell, however this soon gets very complicated and would require the use of delay equations.  Instead we assume that the same amount of product (or service) is made but it costs more or is of lower quality.  For instance, buying in the completed product from some external supplier at market rates to ensure the firm has enough product to sell in the short term.  This would not change the worth of the product to the customer, but would certainly affect the worth to the supplier.  We model this by imposing a multiplicative penalty term ${\Pen}_i$ on the gain in utility given by the selling of product.  This penalty is the actual supply of material and services over the total required supply.  Thus our model for the health/wealth of a firm becomes
\begin{equation}\label{M:Inter}
\dot{u}_i=(1+\epsilon_i) \sum_{j\in C_i}G(u_i,u_j){\Pen}_i-\sum_{j\in S_i} G(u_j,u_i) - d_i u_i.
\end{equation}

The details required to make \eqref{M:Inter} a complete model are expressions for the value of product sold between firms ($G(u_i,u_j)$) and the multiplicative penalty term (${\Pen}_i$).  There are different ways to do this, including full market modelling in the case of the amount of product sold between firms.  However we shall again choose a simpler option.  We shall assume that there is a fixed percentage of the utility that a firm requires in supplies, $\beta_i$.  That is $\beta_i u_i$ is the total amount that firm $i$ must `pay' to buy all of the materials and services it needs when it has grown to size $u_i$.  Similarly we assume that the value of the product each firm creates is proportional to its size $u_i$.  That is a firm with utility $u_i$ produces products with a value $\rho_i u_i$.  A complete list of all the notation used in our model is given in Table~\ref{NOTAT}.  With this notation we can express the penalty term (still for intermediary firms) as
\[{\Pen_i}=\frac{\sum_{j\in S_i} G(u_j,u_i)}{\beta_i u_i},\]
the fraction of the required supply that a firm actually managed to procure.  The expression we use for $G(u_j, u_i)$ is more complicated as firm $i$ is likely to have multiple suppliers and each of them is likely to have multiple customers.  If each firm that supplies firm $i$ can meet all demands for product placed upon them by all of their customers (they're big enough) then firm $i$ is able to buy the total amount of product it requires $\beta_i u_i$.  We assume that the amount it buys from each of its suppliers is proportional to their size (utility).  This expresses the desire to keep the largest of a firm's suppliers as happy as possible - without totally alienating any of the other suppliers.  That is (if all of a firm's suppliers are able to fulfil the entire demand for product placed upon them)
\[G(u_j,u_i)=\beta_i u_i \frac{u_j}{\sum_{k\in S_i} u_k}.\] 

If on the other hand one of firm $i$'s suppliers (firm $j$ say) is unable to fulfil the total demand placed upon it then it will, in total, supply the maximum amount that it can: $\rho_j u_j$.  It will distribute this according to the demand placed on it by each of its customers.  We again assume that it sells in proportion to the amount each customer requires (see above), in order to retain the largest customers without alienating any of the smaller customers (note in our model we do not actually allow the customers or suppliers of firms to change, although they can go out of business).  That is 
\[G(u_j,u_i)=\rho_j u_j \frac{\beta_i u_i \frac{u_j}{\sum_{k\in S_i} u_k}}{\sum_{l\in C_j}\left[\beta_l u_l \frac{u_j}{\sum_{m\in S_l} u_m}\right]}.\]
Combining these two scenarios together gives:
\begin{equation}
G(u_j,u_i)=\left\{\begin{array}{cl}
\beta_i u_i \frac{u_j}{\sum_{k\in S_i} u_k} &,~\mbox{if}~\sum_{l\in C_j}\left[\beta_l u_l \frac{u_j}{\sum_{m\in S_l} u_m}\right]\leq\rho_i u_i,\\
\rho_j u_j \beta_i u_i \frac{u_j}{\sum_{k\in S_i} u_k}\frac{1}{\sum_{l\in C_j}\left[\beta_l u_l \frac{u_j}{\sum_{m\in S_l} u_m}\right]} &,~\mbox{otherwise}.
\end{array}\right.
\end{equation}

\medskip
This completes the details of the model for intermediary firms. We now explain the amendments needed to cope with the `primary suppliers' and `end consumers'.  First of all note that if the current model \eqref{M:Inter} were applied to an end consumer with no customers being modelled then there would be no growth term, $u_i$ would be monotonically decreasing.  Similarly looking at a primary supplier the only negative term would be the overheads, even though products were being bought and paid for (just from outside the set of firms modelled).  To counter this we add an extra term to the model containing the utility gain or loss from exporting or importing, respectively, materials and products from outside the set of firms modelled.  We denote this component of the model $\Lambda_i$.  The value of which depends on the classification of the firm.  If firm $i$ is an intermediary firm then $\Lambda_i=0$.  We assume that the supply of a product from outside the district is essentially unlimited.  That is, a firm which is a supplier can buy what ever it requires from outside of the district, $\beta_i u_i$.  If a firm buys all of its supplies from outside the district then using this makes sense, however if it buys some of its supplies from inside the district it would end up buying supplies to the value of $2\beta_i u_i$.  If this is the case we make the value of $\beta_i$ half what it would otherwise be.  Whilst the choice of a half of the product being bought from within the district and half from outside of the firms modelled is fairly arbitrary, it is something which could easily be improved upon in later iterations of the model if such detailed data is available.

When considering `end consumers' we make the assumption that the external market for the products and services being created is bounded.  The total maximum demand of the external market is for products with value $M$.  If the total amount the district is trying to sell to the external market exceeds the market demand then each end consumer sells according to their size.  
However, in complicated industrial districts it is possible (and indeed likely) that the goods being sold to the external market are not interchangeable.   To account for this we say that there are several external markets for different commodity types, each with a bound on the value of good which can be bought, $M_k$. For instance it seems feasible that fuel could be treated as such a commodity type, with different fuel types competing in the external market due to their similarities.  A further complication arises if a firm supplies multiple distinct external markets, for instance if firm $i$ is in $k_1 \cap k_2$.  In this case, one must know what proportion of the amount sold by firm $i$ goes to market $k_1$ and how much to market $k_2$.  For simplicity we shall assume that firm $i$ wants to sell the same amount of goods in each external market (if in reality the proportion of goods sold to one external market from firm $i$ is much smaller than that sold to other external markets then the sales to that market from firm $i$ could be ignored for the purpose of this model).  In reality a firm with multiple markets may choose to reposition itself if one of those markets becomes unprofitable due to excess competition, however to model this we would need to model the decisions of individual firms, something we choose not to do in order to create a simple deterministic model of the interactions of firms.

If there is just one external market then the amount `end consumer' $i$ sells to the external market is 
\[\tilde{G}(u_i)=\left\{\begin{array}{cl}
\rho_i u_i&, M>\sum_{j \in \bf{C}}\rho_j u_j, \\
M\frac{u_i}{\sum_{j\in \bf{C}}u_j}&, \mbox{ otherwise}.
\end{array}  \right.\]
Again if a firm is an `end consumer' but also sells to other firms in the district then we make $\rho_i$ half of what it would otherwise be.  If firm $i$ is an `end consumer' its utility needs to  increase by $(1+\epsilon_i)\tilde{G}(u_i)$.
That is we have
\[\Lambda_i=\left\{\begin{array}{cl}
-\beta_i u_i&, i \mbox{ a primary supplier}, \\
0&, i\mbox{ an intermediary}, \\
(1+\epsilon_i)\tilde{G}(u_i)&, i\mbox{ an end consumer},\\
-\beta_i u_i+(1+\epsilon_i)\tilde{G}(u_i)&, i \mbox{ a primary supplier and an end consumer}. 
\end{array}  \right.\]
However if there is more than one external market the expression for $\tilde{G}(u_i)$ is slightly more complicated.  First of all we need a rescaled version of $u$, $\tilde{u}$, if firm $i$ sells goods to $n$ external markets then $\tilde{u}_i=u_i/n$.  With this notation 
\[\tilde{G}(u_i)=\sum_{k~|~i\in k}\left\{\begin{array}{cl}
\rho_i \tilde{u}_i&, M_k>\sum_{j \in k}\rho_j \tilde{u}_j, \\
M_k\frac{\tilde{u}_i}{\sum_{j\in k}\tilde{u}_j}&, \mbox{ otherwise}.
\end{array}  \right.\]

Finally we must deal with the multiplicative penalty term due to lack of supply for `primary suppliers' and `end consumers'.  As it is assumed that there is no limit to the amount `primary suppliers' (including hubs) can procure from outside the district, no penalty due to lack of supply will be imposed.  In the case of an `end consumer' (but not a hub) the penalty term will be the same as for the intermediary firms.  However the penalty is applied to the $\Lambda_i$ term as this is now the term which incorporates the benefits to the utility from selling products or services.  That is
\[{\Pen}_i=\left\{\begin{array}{cl}
1 &, i\mbox{ a primary supplier}, \\
\frac{\sum_{j\in S_i}G(u_j,u_i)}{\beta_i u_i} &, \mbox{otherwise}.
\end{array}\right.\]

Thus our model for the metabolic interaction of firms or industries within a local district is  
\begin{equation}\label{M:comp}
\dot{u}_i=(1+\epsilon_i)\sum_{j\in C_i}G(u_i,u_j){\Pen}_i-\sum_{j\in S_i} G(u_j,u_i) +\Lambda_i{\Pen}_i - d_i u_i,
\end{equation}
with the notation detailed in Table~\ref{NOTAT}.
\begin{table}[h!]
\centering
\begin{tabular}{|c||l|}
\multicolumn{1}{c}{\bf Symbol} &  \multicolumn{1}{c}{\bf Definition}\\
\hline 
$\Lambda_a$ & The intrinsic growth rate (seen from a local perspective) of firm $a$.\\ \hline
${\Pen}_i$ & The penalty to firm $i$ of not buying enough product.\\ \hline
$G(a,b)$ & The worth of product $a$ that is bought by firm $b$.\\ \hline
$\beta_a$ & The value of the material or services that $a$ needs as a proportion of its wealth.\\ \hline
$\rho_a$ & The value of $a$'s product or service as a proportion of its wealth.\\ \hline
$\epsilon_a$ & The percentage profit made by firm $a$.\\ \hline
$d_a$ & Percentage upkeep costs of firm $a$.\\ \hline
$M_k$ & The maximum value of product type $k$ which the district can sell externally. \\ \hline
\end{tabular}
\caption{List of Notation}
\label{NOTAT}
\end{table}

\section{A Model of the Humber Region}\label{S:HumberRegion}
We shall use our abstract dynamical model of the interactions of firms or industries to learn more about the intricacies of the Humber region.  As we mentioned in the introduction the Humber region is made up of many distinct industry types.  In~\cite{PJW13} Penn et.~al.~devised a method for creating a complete metabolic network of a system from an incomplete sample of the network using `network archetypes', characteristic sets of input and output connections for given industrial sectors.  They were able to generate a network model of the Humber region by using these archetypes to supplement data from a series of eighteen interviews which were not originally designed to elicit complete network information.   This model gives a static impression of the region and network analysis can be used to draw conclusions about nodes of particular importance or vulnerability wthin the system.  With our dynamical model we seek to go beyond these purely structural results, and learn more about the possible future of the system.  We do not run our dynamical model on Penn et.~al.'s complete model of the Humber region due to the size of the network (seventy-six nodes) and the corresponding difficulty involved in finding the necessary data to initialise our dynamical model.  Instead we use the model of Penn et.~al.~to generate a condensed network representation of the Humber region comprising twenty-two interconnected nodes.  This subset constitutes the the principal bio-based industries in the region and their main connections, direct and indirect, as determined by number of mentions in interview transcripts and local knowledge on magnitude of material supplies. Importantly this network includes every stage in the possible bio-based economy cycle from primary producers (agriculture), via processors of different kinds (food, biofuels etc) to waste processors and recyclers (landfill, anaerobic digestion, composters etc).

\begin{figure}[!h]
\centering
\includegraphics[width=6in]{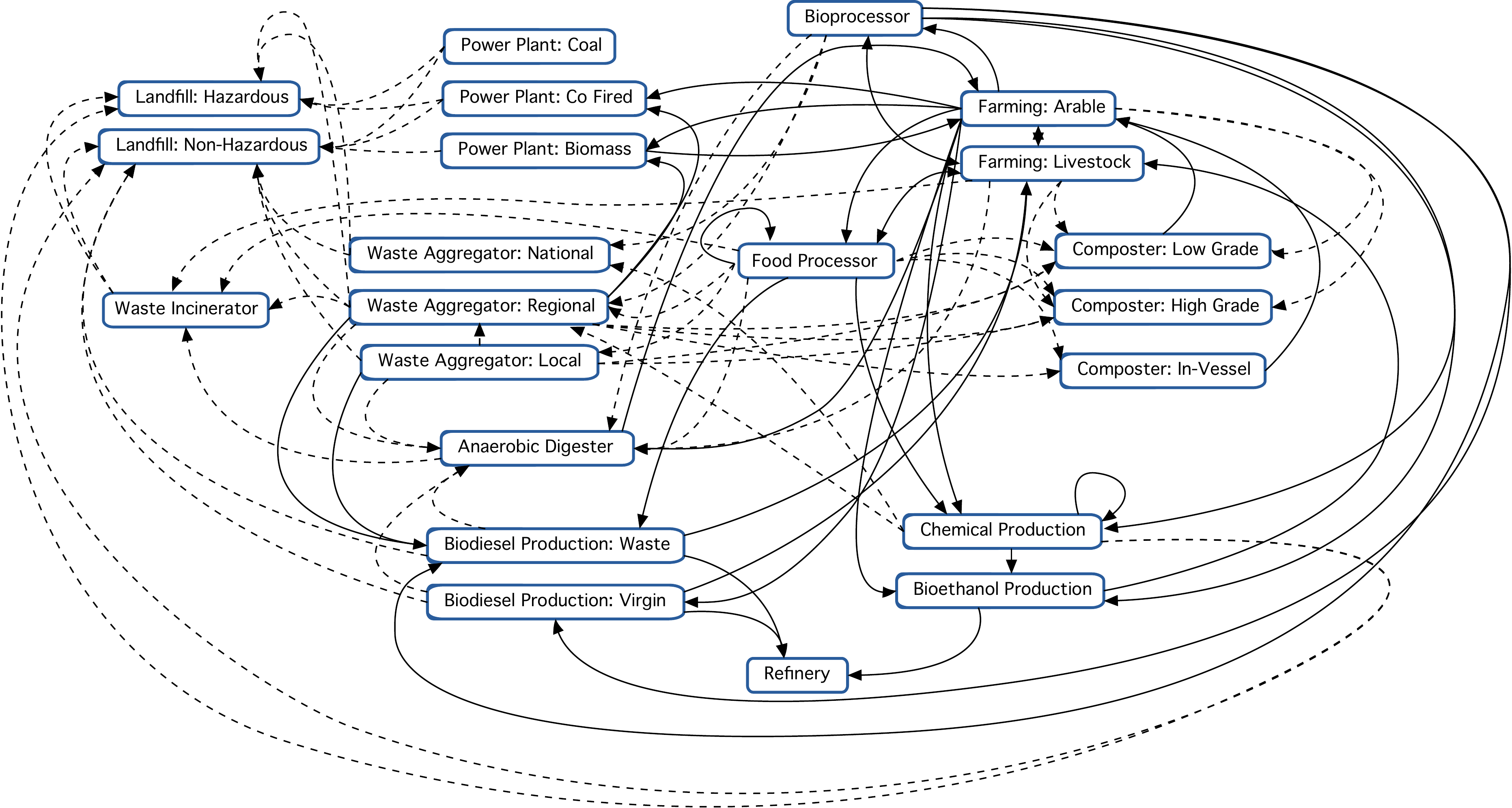}
\caption{A condensed model of the industry interactions of the Humber region.  Arrows show the flow of material.  Solid lines depict instances in which the flow of money is in the opposite direction to the flow of material, whilst dashed lines represent instances where money and material flow in the same direction (e.g. waste disposal).}
\label{F:HumberNetwork}
\end{figure}

Figure~\ref{F:HumberNetwork} shows the network model for the metabolic interactions of the industries in the Humber region.  
The edges represent flow of material whilst the style of the edges represent how the flow of material relates to the flow of money.  In building our dynamical model of firm interaction we have implicitly assumed that the flow of material is opposite to the flow of money; that is you buy a material or product.  These are represented by solid lines in the figure.  However in the Humber region there are several industries which deal with waste (for instance landfill sites), these industries are paid to take the waste from other industries and dispose of it.  In this case the material and the money both move in the same direction.  These are the dashed lines in Figure~\ref{F:HumberNetwork}.   To allow the use of our dynamical model~\eqref{M:comp} on this representation of the region we must find a representation where the flow of money is opposite to the flow of material or service.  This is the key point, whilst the material may flow in the same direction as the money this in fact represents a service flowing in the opposite direction.  For instance, waste flows into the landfill, however if instead of the waste we consider the service that the landfill is providing in taking waste from other industries then this flows in the opposite direction to the flow of money.  Thus the network that we actually use to represent the Humber region in our dynamical model has all of the dashed arrows reversed.  So, for instance, landfill has lots of out going edges and no incoming edges, making it a primary supplier.  The classification of the other nodes in terms of their role with relation to firms outside the district is shown in Table~\ref{T:class}.
\begin{table}[h!]
\centering
\begin{tabular}{|l||c|c|}
\multicolumn{1}{c}{\bf Industry Type} &  \multicolumn{1}{c}{\bf Category} & \multicolumn{1}{c}{\bf External Product}\\
\hline 
Refinery: Crude Oil & Hub & Fuel, Chemicals\\ \hline
Food Processor & End Consumer & Food\\ \hline
Farming: Arable & Primary Supplier &\\ \hline
Farming: Livestock & Primary Supplier &\\ \hline
Power Plant: Coal-Fired & Hub & Electricity\\ \hline
Power Plant: Co-Fired & Hub & Electricity\\ \hline
Composter: Low Grade & Primary Supplier &\\ \hline
Composter: High Grade & Primary Supplier &\\ \hline
Composter: In-Vessel & Primary Supplier &\\ \hline
Waste Incinerator & Hub & Electricity\\ \hline
Landfill: Non-Hazardous & Primary Supplier &\\ \hline
Landfill: Hazardous & Primary Supplier &\\ \hline
Waste Aggregator: National & Intermediary &\\ \hline
Waste Aggregator: Large Regional & Intermediary &\\ \hline 
Waste Aggregator: Small Local & Intermediary &\\ \hline  
Power Plant: Biomass & Hub & Electricity\\ \hline 
Biodiesel Production: Virgin Feedstock & Primary Supplier &\\ \hline 
Anaerobic Digester & End Consumer & Electricity\\ \hline 
Chemical Production: Biological & Hub & Chemicals\\ \hline 
Bioethanol Production: Virgin Feedstock & Hub & Fuel\\ \hline 
Biodiesel Production: From Waste & Hub & Fuel\\ \hline 
Bioprocessor & Intermediary &\\ \hline 
\end{tabular}
\caption{Industry Classifications and External Market Products.}
\label{T:class}
\end{table}

In order to run the dynamical model~\eqref{M:comp} for industrial interactions on the network of the Humber region we first need to find initial values for the health/wealth ($u_i$), as well as parameters; $\beta_i$, $\rho_i$, $\epsilon_i$ and $d_i$, for each of the industry types.  We also need to classify the product types that are sold to the external market by the end consumers and hubs.  The major commodities which the firms in our model export are; fuel, chemicals, electricity, and food.  The external markets that each firm supplies can be found in Table~\ref{T:class}.  For these four different external markets we must find values for the maximal demand, $M_k$.  

Due to the fact that not all firms which would be classified as the same industry type operate in the same way, finding exact values for all of the parameters and initial conditions is not possible.  For instance consider arable farming - the economics will depend on which crop is grown as well as the size of the farm, and what method is used to grow the crop.  Whilst the price the crop fetches will depend on the quality of the crop (which depends on the weather and on disease spread) and how many other farms have grown the same crop.  Instead we use average values, or values derived from an `archetype' firm of a particular industry.  The values that we use for the parameters and initial conditions are given in Table~\ref{T:IC} and their derivation can be found in Appendix~\ref{A:data}.  We take the total external market demand $M$ to be the sum of the initial utilities over all end consumers, that is $M=13777.42$.   For the individual markets, e.g.~fuel, we take the sum of the initial rescaled utilities, $\tilde{u}_i$ over all firms supplying that market.  So for instance the external market demand for chemicals $M_{C}$ is the initial utility of ``chemical production: biological" plus half the initial utility of the refinery, $M_C=6573.33$.  The external market demand for fuel is $M_{Fl}=4870.58$, the external market demand for electricity is $M_E=2308.54$ and the external market demand for food is $M_{Fd}=24.99$.

Instead of viewing all of these numbers for the parameters as exact we must consider them as only showing the relative magnitude of the different parameters.  This means that any analysis based on the results of the dynamical model must be treated carefully and not blithely stated as guaranteed predictions of the future.
\begin{table}[h!]
\centering
\begin{tabular}{|l||c|c|c|c|c|}
\multicolumn{1}{c}{\bf Industry Type}  &  \multicolumn{1}{c}{\bf $u(0)$}  &  \multicolumn{1}{c}{\bf $\beta$}  &  \multicolumn{1}{c}{\bf $\rho$}  &  \multicolumn{1}{c}{\bf $\epsilon$}  &  \multicolumn{1}{c}{\bf $d$}\\
\hline 
Refinery: Crude Oil & 9445.45	& 0.228 & 0.493 & 0.027 & 0.007\\ \hline
Food Processor & 24.99	 &	0.153 & 0.238 & 0.119 & 0.191\\ \hline
Farming: Arable & 0.11	&	0.061	&	0.444 &	0.507 &	0.066\\ \hline
Farming: Livestock & 0.28	&	0.105	&	0.465 &	0.184	 & 0.029\\ \hline
Power Plant: Coal-Fired & 451.62	&	0.080	&	0.226 &	0.690	 & 0.046 \\ \hline
Power Plant: Co-Fired & 1627.35	 &	0.080	&	0.227 & 	0.714 &	0.038 \\ \hline
Composter: Low Grade & 0.25	&	0	&	0.470 &	0.845 &	0.045 \\ \hline
Composter: High Grade & 0.25	&	0	&	0.470 &	0.845 &	0.045 \\ \hline
Composter: In-Vessel & 0.25	&	0	&	0.470 &	0.845 &	0.045 \\ \hline
Waste Incinerator & 11.76	&	0	&	0.234 & 0.163 &	0.104\\ \hline
Landfill: Non-Hazardous & 4.97	&	0	&	0.444 &	0.484 &	0.140 \\ \hline
Landfill: Hazardous & 2.55	&	0	&	0.444 &	0.502 &	0.136 \\ \hline
Waste Aggregator: National & 1590.72 &	0	&	0.497 & 0.011 &	0.032\\ \hline
Waste Aggregator: Large Regional & 159.07 &	0	&	0.497 & 0.011 &	0.032\\ \hline 
Waste Aggregator: Small Local & 1.59 &	0	&	0.497 & 0.011 &	0.032 \\ \hline  
Power Plant: Biomass & 209.21	&	0.167	&	0.239	& 0.106 &	0.030\\ \hline 
Biodiesel Production: Virgin Feedstock & 37.57	&	0.138	&	0.484 &	0.073 &	0.048\\ \hline 
Anaerobic Digester & 8.59	&	0	&	0.225 &	0.661 &	0.013 \\ \hline 
Chemical Production: Biological & 1850.60	&	0.173	&	0.229	& 0.241 &	0.056 \\ \hline 
Bioethanol Production: Virgin Feedstock & 120.87	&	0.084	&	0.223 &	0.570 &	0.019 \\ \hline 
Biodiesel Production: From Waste & 26.98	&	0.076	&	0.225	& 0.340 &	0.067 \\ \hline 
Bioprocessor & 199.61	 &	0.383	&	0.470 &	0.148 &	0.014 \\ \hline 
\end{tabular}
\caption{The initial conditions and parameters needed in the dynamical model of the Humber region~\eqref{M:comp}.  A detailed description of the method used to derive these values and the sources used can be found in Appendix~\ref{A:data}. }
\label{T:IC}
\end{table}

\subsection{Temporal Changes in the Humber Region}
\begin{figure}[!h]
\centering
\includegraphics[width=5.7in]{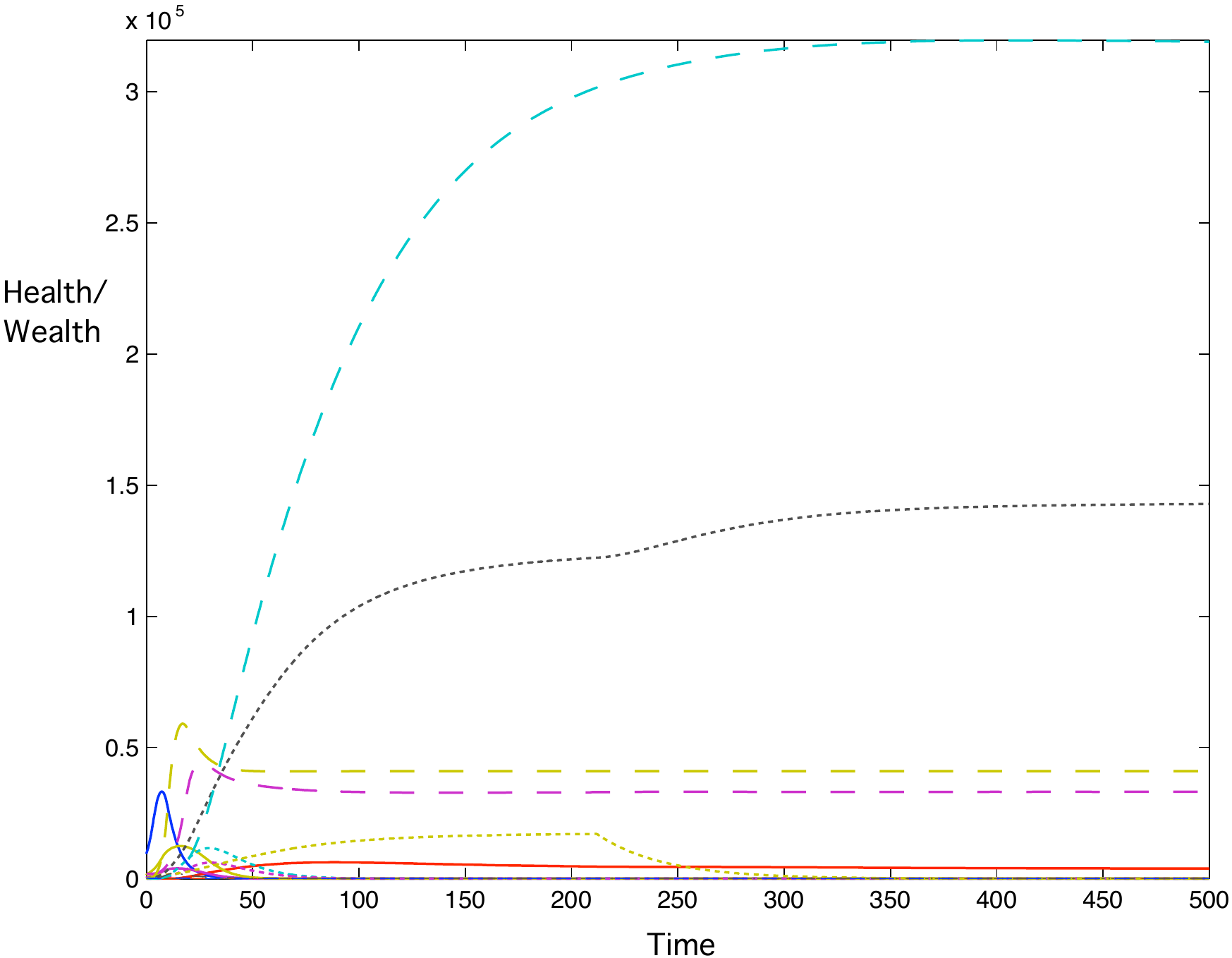}
\includegraphics[width=5.7in]{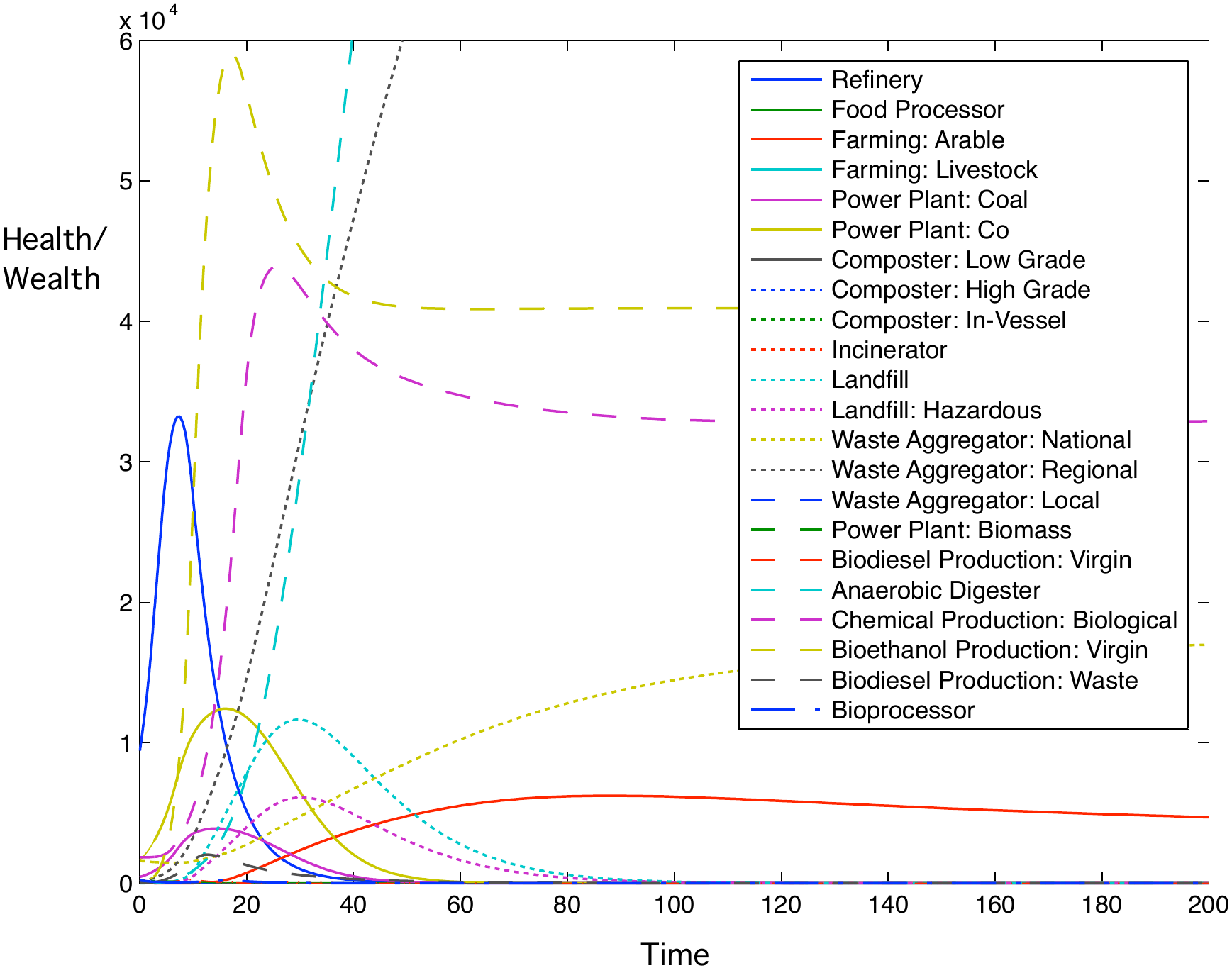}
\caption{How the utility of various industry types changes over time according to our model for the interactions of industries~\eqref{M:comp} on the model of the structure of the Humber region, Figure~\ref{F:HumberNetwork}, with the parameters as listed in Table~\ref{T:IC}.}\label{F:ModelOutput}
\end{figure}

In this section we present the results that our dynamical model of firm (or industry type) interaction~\eqref{M:comp} gives when applied to the network model of the Humber region shown in Figure~\ref{F:HumberNetwork} with parameters initialised as shown in Table~\ref{T:IC}.  To do this we have written matlab code to run the system of differential equations~\eqref{M:comp} using the built in ode45 procedure to iterate the initial conditions.  The output is shown in Figure~\ref{F:ModelOutput}, which plots how the utility of the different industry types changes over time.
Due to the difference in scales between the maximum utilities of the various industry types we show a zoomed in version as well, enabling a more detailed view of the behaviour of several industries to be seen.

We point out that the dynamics of the various utilities are unlikely to be accurate for long time scales due to the assumptions that were made in creating the dynamical model.  For instance we assume that all products (sold locally) are interchangeable, which allows industries to die off entirely.  This wouldn't happen to the same extent in reality as many  goods are non-interchangeable.  Another issue is the fact that new connections are not made, meaning that if all suppliers go out of business then so will the dependent industry.  All of this means that the dynamics of the utilities of the various industries shown in Figure~\ref{F:ModelOutput} are only valid for short time scales, and in particular are unlikely to be valid much beyond when the first industry fails.  From the output we shall take this point to be around time step 40 (nominally 40 years although the real time period is likely to be shorter).  Thus we analyse the output (in narrative form) of the model up until this point.

Initially the utility of almost every industry type increases suggesting that the structure of the industries in the Humber region is viable, at least in the short term.  The only industry types which show a decrease in utility initially are the national waste aggregator, the bioprocessor, and the biomass power plant, all of which only decrease very slowly.  Initially the utility of the food processor is almost static and only changes slowly over time, this is due to the fact that it is the only industry type in the district which supplies the external food market - a market (from the district perspective) which is substantially smaller than all of the other external markets.  This substantially constrains the food processor meaning it will remain smaller than many other of the industry types in the district. 

The first interesting transition appears to occur at time 7, when the refinery reaches a peak and starts to decline drastically.  This is caused by the limited external market for fuel and chemicals.  On the chemical side the refinery is being replaced by chemical production from biological feedstocks, whilst on the fuel side it is being replaced by biodiesel production (primarialy from waste) and bioethanol production from virgin feedstock.  It is particularly interesting to see the transition from an oil based economy to one centred on bioethanol (from virgin feedstock) and biodiesel (from waste) production in such a short time-scale as the Humber region is currently trying to shift its economy in this direction, away from an oil-centric economy towards a bio-based economy, see~\cite{PKL13} and references therein.  However, in the slightly longer term we see that biodiesel production starts decreasing, again being replaced by bioethanol production.  In a similar time frame as this decrease in biodiesel production the model suggests that the district will have over estimated the demand of the external market for both fuel and chemicals, resulting in a sharp decrease in their production until an equilibrium is reached.

Another transition that we can see in Figure~\ref{F:ModelOutput} is in the source of electricity.  After the initial increase in all forms of electricity generation (with the exception of biomass power plants), both coal-firing and co-firing power plants start to decrease with electricity instead being produced by anaerobic digestion.  The reason such a transition appears possible is that there is an increase in the amount of waste in the system, with landfills and waste aggregators growing, eventually landfills start decreasing again with the waste instead going to the waste aggregators (both national and regional).
As anaerobic digestion increases it takes more and more waste away from landfill, effectively becoming the dominant form of waste disposal in the region.  The increase in anaerobic digestion leads to an increase in the amount of digestate that is available for farmers (arable) to use, meaning that they can grow their crops more cheaply and thus  leading to an increase in their utility.   There is also a feedback effect here, in that an increase in arable farming leads to more goods which can be used in the anaerobic digestion process. 

In summary it appears that in the short term the dominant industry of the Humber region will transition from refineries to bio-ethanol production (from virgin feedstock) and biological chemical production.  As the production of electricity changes these industries are replaced in their dominance in the district by anaerobic digestion, and the associated regional waste aggeregator. 
\section{Conclusion}
In this paper we have developed a simple dynamical model describing how different firms or industries interact via their supply chains.  The combination of the supply chains of various industries forms an interconnected web of such metabolic interactions.  The model that we have developed acts on this network of metabolic connections, following the impact of an increase or decrease of an industry's wealth/health on other industries and their consequential impact on each other.

This dynamical model could be used in many different scenarios to explore how different industrial districts or regions might function, and to investigate the possible future of the district.  This could be especially useful if a major transition in focus of the district is occurring, or is expected to occur in the near future.  We applied our model to the Humber region in the UK, a major source of UK energy production traditionally from fossil fuels and also of oil refining capacity.  The region is currently undergoing a transition towards a bio-based economy.  Using a (roughly) parametrised version of the dynamical model we were indeed able to see the development and transition to a bio-based economy.  The outcomes of the model seem to make sense within the wider context of the Humber region and are what is hoped for in a transition from fossil fuel based economy to a bio-based economy.  The oil refinery is replaced (in dominance) by bioethanol production from virgin feedstock, biodiesel production from waste, and chemical production from biological feedstocks.  On the power generation side, the use of both coal and co-firing power plants increase before being replaced by electricity generated from waste through anaerobic digestion.  Such an increase in anaerobic digestion may well require more waste material than can be provided easily, which could lead to an increase in the use of raw materials from arable farmers.  However the excess digestate such anaerobic digestion produces would enable cheaper enrichment of soil for arable farming.  The fact that these outcomes seem to be desirable suggests that the current structure and make-up of the Humber region is adequate to the political desire to transition to a bio-based economy.

However, even though the the dynamical output of our model makes sense we must still sound a note of caution.  The dynamical model assumes that all goods are interchangeable (locally) and the parameter values used are only rough estimates.  There are also major aspects of the industrial district that are not modelled at all.  The most obvious (and probably influential) of which is the social aspect of the district.  It is simply not true that the only way that firms or industries influence each other is through supply and demand metabolic interactions.  There are many different forms of social interaction, such as joint bidding for funding, competition for scarce highly skilled workers, and joint training. Each of these forms of social relationship potentially detailing its own network of relations between the firms or industries in a district.  It does, however, seem possible to divide the social relationships into two broad classifications; those which are mutually beneficial, and those which are detrimental in some way to the firms involved.  We explore this issue further in~\cite{KPH13}, extending the dynamical model presented here to incorporate these social interactions.

\section*{Acknowledgements}
This work was supported by the Engineering and Physical Sciences Research Council [grant number EP/H021450/1] (Evolution and Resilience of Industrial Ecosystems ERIE). We would like to thank D.~Avitabile for useful discussions during our initial model development, and F.~Schiller, A.~Yang and E.~York for their help in finding some of the sources used in parametrising the model for the Humber region.

\pagebreak
\appendix
\section{Derivation of Parameter Values Used.}\label{A:data}
This section contains the derivation of the parameter values found in Table~\ref{T:IC}, along with the sources of the data used in the derivation.  As there are 111 separate parameter values which need evidencing it should not be a surprise that we used a multitude of sources of data.  As we only require rough numbers for the parameters many of these sources are web pages.  It was fairly easy to find data for some of the industry types, where for instance we could access full accounts for an archetype company, however in other instances it was harder or impossible to find the specific data required.  For instance we ended up using the same data for all three types of composter, and the regional and local waste aggregators were based on national waste aggregators with an attempt at appropriate scaling applied.  With all of these issues, along with the fact that some data sources had to be inflation and currency adjusted, we must emphasise that the data in Table~\ref{T:IC} should only be viewed as giving approximate orders of magnitude for the parameters, and any conclusions drawn should be carefully analysed.

In order to find the necessary parameter values we found four pieces of data for each industry type in 2013 GBP: the total revenue in a year, the amount spent on materials in a year, the amount spent on overheads in a year, and the amount spent on production in the year, see Table~\ref{T:Data}.  (Note that the amount spent on material in a year can be negative if the industry is paid to take that material, for instance landfill.)  
\begin{table}[h!]
\centering
\hspace*{-0.7in}
\begin{tabular}{|l||c|c|c|c|}
\multicolumn{1}{c}{\bf Industry Type}  &  \multicolumn{1}{c}{\bf Revenue}  &  \multicolumn{1}{c}{\bf Material Cost}  &  \multicolumn{1}{c}{\bf Overheads}  &  \multicolumn{1}{c}{\bf Production Cost} \\
\hline 
Refinery: Crude Oil & 4787.96	& 4311.51 &	69.2	& 276.78\\ \hline
Food Processor & 13.28 &	3.826	& 4.76	& 3.12\\ \hline
Farming: Arable & 0.071	& 0.013 &	0.007	& 0.015\\ \hline
Farming: Livestock & 0.152	& 0.058	& 0.008	& 0.058\\ \hline
Power Plant: Coal-Fired & 344.7	& 72.67	& 20.55	& 13.7\\ \hline
Power Plant: Co-Fired & 1265.04	 & 259.56 & 	61.65	& 41.1 \\ \hline
Composter: Low Grade &  0.151 &	-0.062	& 0.011	& 0.022\\ \hline
Composter: High Grade &   0.151 &	-0.062	& 0.011	& 0.022\\ \hline
Composter: In-Vessel &   0.151 &	-0.062	& 0.011	& 0.022\\ \hline
Waste Incinerator & 4.106	& -2.296	& 1.22	& 4.14\\ \hline
Landfill: Non-Hazardous &  0	& -3.28	& 0.695	& 0.998\\ \hline
Landfill: Hazardous &  0	& -1.7	& 0.3475	&0.499\\ \hline
Waste Aggregator: National & 143.55	& -656.35	& 51.08	& 739.74\\ \hline
Waste Aggregator: Large Regional & 14.355	& -65.635	& 5.108	& 73.974\\ \hline 
Waste Aggregator: Small Local &  0.14355	& -0.65635	& 0.05108	& 0.73974\\ \hline  
Power Plant: Biomass & 110.48	& 70	& 6.23	&22.5\\ \hline 
Biodiesel Production: Virgin Feedstock & 19.5	 &10.37	& 1.8	& 5.9\\ \hline 
Anaerobic Digester &  4.37	& -2.05	& 0.114	& 2.06\\ \hline 
Chemical Production: Biological &  1051.9	& 638.9	& 104.1	& 55.7\\ \hline 
Bioethanol Production: Virgin Feedstock &  84.5 & 	20.33	& 2.24	& 13.8\\ \hline 
Biodiesel Production: From Waste & 16.25	& 4.11	& 1.8	& 4.82 \\ \hline 
Bioprocessor &  107.76	& 76.5	 & 2.78 &	12.57\\ \hline 
\end{tabular}
\caption{The data used to derive the initial conditions and parameters needed in the dynamical model of the Humber region~\eqref{M:comp}. Values are millions of 2013 GBP.}
\label{T:Data}
\end{table}
We used these pieces of data to derive the parameters used in the modes.  For the wealth/health of a firm ($u_i$) we used the sum of the revenue, the absolute value of the material cost, the overhead cost and the production cost.  For the percentage spent on material ($\beta$) we used $0$ if the total material cost was zero, otherwise we used the cost of materials over $u_i$.  This was further divided by two if the industry was a primary supplier and bought material from within the district.  For the percentage profit ($\epsilon$) we used the value of goods and services (revenue + $|$material cost$|$ if material cost is negative, or revenue otherwise) less the cost of material bought (zero if total material cost was negative) less overheads less production costs all over the value of goods and services.  For $\rho$ we used the value of goods and services over $u_i(1+\epsilon_i)$.  This is further divided by two if the industry is an end consumer and also sells to firms in the district.  Note that the $\rho$ term has been divided by $(1+\epsilon_i)$ to recover the actual `worth' of the product before profit was included.  Finally for the total external demand of the market ($M$) we use the sum over all end consumers of $u_i$.

We now give separate details of the data sources for each of the industry types.  For currency conversion we use the following factors throughout: \$1=\pounds 0.65, \pounds 1=\euro 1.25. 
\subsection{Refinery: Crude Oil}
We base our analysis for a refinery on the Total Lindsay oil refinery~\cite{Refinery1} which produces 10 million tonnes per year, that is 73.3 million barrels a year,~\cite{Refinery2}.  We use spot prices to find the average price of crude oil~\cite{Refinery3}, between 2007 and 2012 these spot prices were (USD/barrel); 80.14, 104.84, 66.19, 83.76, 105.20 and 102.84.  This gives an average cost per barrel of \$90.50 or \pounds 58.82.  Thus the cost of materials is $58.82\times 73.3 M=4,311.51M$.

The gross margin (that is sales - cost of materials) per barrel is given in \cite[Table 3]{Refinery4}.  They give average gross margins for 2001 to 2005 (USD/barrel); 8.60, 6.89, 8.36, 8.05 and 9.87.  Thus the average gross margin is \$8.35/barrel, or equivalently \pounds 6.50 (inflation adjusted).  So the total revenue is $(58.82+6.5)\times 73.3M = 4,787.96 M$.  \cite[Table 3]{Refinery4} also gives the average net margins (gross margins less overheads less production costs).  From 2001 to 2005 they were (USD/barrel); 2.99, 0.21 2.18, 2.56 and 3.51, which gives and average net margin of \$2.29 or \pounds 1.78 (inflation adjusted per barrel).  Thus the cost of overheads and production (per barrel) is $\pounds 6.50-\pounds 1.78=\pounds 4.72$.   Now, \cite[p.~101]{Refinery5}, \emph{``Fixed costs can represent up to 80\% of the total cost of processing every tonne of crude"}.  We shall use the fixed costs as the production costs, that is \pounds 3.776/barrel, so the overheads are \pounds 0.944/barrel.  Multiplying both these figures by $73.3M$ gives the values presented in Table~\ref{T:Data}.

\subsection{Food Processor}
We base our analysis of food processors on Cadbury, as suitable data was available, and then scale the resulting figures to better represent an average food processor.  We use data from the 2009 annual financial report as this is the last financial report before Cadbury was bought out and amalgamated by Kraft Foods.  Specifically the data is from stock exchange RNS announcement 2009 (annual financial report)~\cite{FP1}.  This document reports Revenue (sales) of $\pounds5,384 M$, underlying profits of $\pounds638 M$, trade payable of $\pounds1,551 M$, distribution cost of $\pounds 247 M$, marketing costs of $\pounds 584 M$, and administrative costs of $\pounds1,098 M$.

We use the trade payable figure as the cost of materials, and the combination of distribution, marketing and administrative costs for the overheads ($\pounds1,929 M$).  For the production costs we use the sales less profits less materials less overheads ($\pounds1,266 M$).  Finally we need to scale these figures to represent a more average food processor.   The average firm revenue in the food processing sector is $\pounds 10.5 M$ ($\pounds 13.28 M$ inflation adjusted)~\cite[Table 2.1]{FP2}.  Dividing the figures from Cadbury by $5384/13.28\approx 405.4$ gives the values in Table~\ref{T:Data}.

\subsection{Farming: Arable}
There are many different arable crops grown in the UK, however the vast majority of the crop produces is cereals, and of this the largest crop is wheat~\cite{AF1}.  As such we shall base our analysis of arable farming on growing wheat.  We also use the fact that the average UK farm size is 57 hectares~\cite{AF2}.  From~\cite[Table 4.4]{Extra4} we learn approximate costs and revenue per hectare;  the returns (sales) are $\pounds 1,110/ha$ ($\pounds 1,254$ inflation adjusted), the cost of agrochemicals is $\pounds 143/ha$ ($\pounds 161.58$ inflation adjusted), and total variable costs of $\pounds 551/ha$ ($\pounds 622.57$ inflation adjusted).

In order to provide an estimate for the amount spent on materials we shall use the amount spent on agrochemicals and the amount spent on seed. \cite{AF3} gives the cost of seed as $\pounds 50/ha$ ($\pounds 73.47$ inflation adjusted), which gives a materials cost of $\pounds 235.05/ha$.  For the overheads we shall use the fuel costs,~\cite[Table 4]{Extra5} gives the amount of different fuels that are used in growing cereals per hectare. We use spot prices (taken in mid September 2013) for these fuels to calculate a total fuel cost per hectare of $\pounds 120.71$, see Table~\ref{T:AFfuel}.
\begin{table}[h!]
\centering
\begin{tabular}{|l||c|c|}
\multicolumn{1}{c}{\bf Fuel Type}  &  \multicolumn{1}{c}{\bf Amount (per ha.)}  &  \multicolumn{1}{c}{\bf Unit Cost} \\
\hline 
Road fuel & 11 L & 142.47p\\ \hline
Red Diesel & 115.6 L &  69.5p\\ \hline
LPG & 2 Kg ($\approx$4 L) & 70.12p (/L)\\ \hline
Kerosene & 11.8 L & 58.3p\\ \hline
Electricity & 115.5 KWh & 13p\\ \hline
\end{tabular}
\caption{Fuel use per hectare and unit cost for growing cereals.}
\label{T:AFfuel}
\end{table}

Thus with total variable costs per hectare of $\pounds 622.57$, material costs of $\pounds 235.05$, and overheads of $\pounds 120.71$, the production costs must be the difference (variable costs less material costs less overheads) which gives a figure of $\pounds 266.81/ha$.  Multiplying all of these figures by the 57 hectare average farm size gives the values used in Table~\ref{T:Data}.

\subsection{Farming: Livestock}
Like arable farming, livestock farming is diverse in the UK, however unlike arable farming it is not dominated by one crop.  We choose to focus on dairy farming and base our analysis on that as~\cite{Extra5} also provides data on fuel use for dairy farming, giving a total fuel cost (per hectare) of $\pounds192.52$, see Table~\ref{T:LFfuel}.  As for arable farming we use this fuel cost as the overheads.
\begin{table}[h!]
\centering
\begin{tabular}{|l||c|c|}
\multicolumn{1}{c}{\bf Fuel Type}  &  \multicolumn{1}{c}{\bf Amount (per ha.)}  &  \multicolumn{1}{c}{\bf Unit Cost} \\
\hline 
Road fuel & 13.7 L & 142.47p\\ \hline
Red Diesel & 147.9 L &  69.5p\\ \hline
LPG & 0.7 Kg ($\approx$1.4 L) & 70.12p (/L)\\ \hline
Kerosene & 2.9 L & 58.3p\\ \hline
Electricity & 511.8 KWh & 13p\\ \hline
\end{tabular}
\caption{Fuel use per hectare and unit cost for dairy farming.}
\label{T:LFfuel}
\end{table}
The average dairy farm in the UK has a herd size of 86 cows, and an average density of two cows per hectare~\cite{AF1}, which makes the average dairy farm 43 hectares, allowing the total fuel use to be calculated. 

\cite{AF1} also states that the average dairy cow yields $6,300$ litres of milk a year.  With a spot price for milk at farm gate (taken in mid September 2013) of $28.04p/L$ we have an average dairy farm revenue of $6300\times 86 \times 0.2804 = \pounds 151,921$.  
Finally~\cite{LF1} gives a figure for the total production costs of $21.5 p/L$ whilst~\cite{LF2} states \emph{"feed costs represent[] 40 to 60 percent of the cost of producing milk".}  We take the average of 50\% of total production costs being spent on feed (materials), meaning that material costs and the production costs (excluding materials) in Table~\ref{T:Data} are both same; $0.215\times0.5\times 6300\times 86=\pounds 58,243.5$. 

\subsection{Anaerobic Digester} 
We base our analysis of Anaerobic digestion on GWE Biogass Ltd which converts 50,000 tonnes of waste a year to biogas which is the converted to 2MW of electricity~\cite{AD1}.  If we assume that the plant operates for the equivalent of 300 complete days a year it generates 14.4 GWh of electricity.  Further~\cite{AD2} gives that the mass of sludge (digestate) out is 80\% to 90\% of the mass in.  Using the figure of 80\% gives that the plant produces $40,000$ tonnes of sludge a year.  Schiller~\cite{PC:FS} suggests that digestate has a value of $\pounds 50/tonne$, giving a revenue from digestate of $\pounds 2 M$.  From~\cite{Extra3} we can calculate the price the firm gets from sale of electricity, Table A-4 gives a basic price of \euro{}$58.02/MWh$ and Table A-8 gives an additional price for electricity support (to incentivise production of electricity from `green' sources) of \euro{}$123.36/MWh$.  This gives a total gain of \euro{}$181.36/MWh$ ($\pounds 164.78/MWh$ inflation adjusted).  Meaning that the revenue from selling electricity is $\pounds 2.37 M$ per year.  So the total revenue (sales) is $\pounds 4.37 M$. 

The price for materials is negative.  The anaerobic digester uses waste from other industries as its source material.  If it didn't use this material then the industry would have to pay for its disposal, (traditionally through landfill or incineration).  Schiller~\cite{PC:FS} suggests a price of $\pounds 41/tonne$ to dispose of waste.  As GWE biogas use 50,000 tonnes of waste a year their material costs are $-\pounds 2.05 M$.

Finally the overheads and production costs. \cite[Table 4]{AD3} gives the overheads for an anaerobic digestion plant with 70,000 tonne capacity as \euro{}$175K$ ($\pounds 158.98 K$ inflation adjusted).  A simple scaling for the difference in capacity gives an approximation of the overheads for GWE biogas as $\pounds 113.56 K$.  For the production costs we use the OPEX costs as given in~\cite[p.~36]{Extra3}, a value of \euro$45.25/tonne$ ($\pounds 41.11/tonne$ inflation adjusted).  Multiplying this by the 50,000 tonne capacity gives the value used in Table~\ref{T:Data}.

\subsection{Power Plant: Coal-Fired}\label{A:PPC}
We base our analysis of coal-fired power plants on a 900MW capacity plant running for 7880 hours (full time equivalent) a year.  Our analysis is based on these specifics as this forms one of the cases studied in~\cite{PPC1}.  In \cite{PPC1} different technologies are considered as well as different coal types.  We shall average the data over these technological and coal type differences to arrive at the values used.  The technologies used are described as subcritical, supercritical and ultra supercritical, whilst the three coal types analysed are Bituminous, Lignite and PRB. 

Now,~\cite[Appendix A]{PPC1} gives the annual production of electricity (in TWh) across these 9 scenarios as 6.53, 6.47, 6.60, 6.54, 6.48, 6.61, 6.54, 6.49 and  6.61.  Thus the average electricity generation is $6.54 TWh$.  We assume that this electricity is sold at the same basic price (no subsidies) as that generated by anaerobic digestion, i.e.~\euro{}$58.02/MWh$ ($\pounds 52.71/MWh$ inflation adjusted), this gives the sales figure for coal-fired power plants.

In order to calculate the material cost, we need to know how much coal is burned and the cost of that coal.  The cost of coal to the gate is given in \cite[Appendix A]{PPC1} as \$39.55, \$17.90 and \$23.47 per tonne depending on type and quality.  This gives an average price of \$$26.97/tonne$ ($\pounds 19.05/tonne$ inflation adjusted).  Further~\cite{PPC2} says that a 100MW power plant uses 53.8 Tonnes of coal an hour.  Assuming no efficiencies of scale this means that a 900MW power plant uses 484.2 tonnes of coal an hour.  The annual material cost is thus given by $\pounds 19.05 \times 484.2 \times 7880 =\pounds 72.67 M$.

\cite[Appendix A]{PPC1} also gives values for fixed and variable costs.  We use the variable costs for the overheads and the fixed costs for production costs.  So for overheads we take the average of 27.33, 38.51, 20.03, 27.64, 38.99, 20.24, 28.35, 40.06 and 20.71 (\$M), which is \$$29.1 M$ ($\pounds 20.55 M$ inflation adjusted).  For the production costs we take the average of 19.07, 19.43 and 19.61 (\$M), which is \$$19.4 M$ ($\pounds 13.70 M$ inflation adjusted).

\subsection{Power Plant: Biomass}
We base our analysis of a biomass-fired power plant on the proposed Drax Heron plant, which would burn 1.4 million tonnes of biomass a year to produce 290MW of electricity~\cite{PPB1}.  Again to calculate the revenue we shall use the same basic price for electricity generated as for the anaerobic digester, namely \euro{}$58.02/MWh$ ($\pounds 52.71/MWh$ inflation adjusted).  There used to be an additional subsidy linked to how much biomass was burnt~\cite{PPB2}, but that has recently been repealed.  Thus, assuming that the power plant operates for 7200 (full time equivalent) hours a year the revenue from electricity generation is $\pounds 110.48 M$.  To calculate the material cost we just need a price for biomass.  We use the value for wood chips of $\pounds 50/tonne$ given in~\cite{PPB3}, meaning that the total material cost is $\pounds 70 M$.  

To calculate the production costs we note that~\cite{PPB3} gives annual operating costs of 2\% of capital costs and annual maintenance costs of 2.5\% of capital costs.  The 250MW power plant that they use in their analysis had a capital cost of $\pounds 400 M$.  Scaling with no economies of scale gives a capital cost for a 290MW plant of $\pounds 464M$ ($\pounds 524 M$ inflation adjusted).  However it is likely that economies of scale should be taken in to effect, as a result we shall use a capital cost of $\pounds 500M$.  We use the annual operating and maintenance costs as a proxy for the production costs giving a production cost of $\pounds 22.5 M$.   Finally, we need a value for the overheads.  We were unable to find specific data for biomass burning power plants, so instead we use data for coal-burning power plants given in~\cite[Appendix A]{PPC1}.   As for the coal-firing power plant we use the total variable costs averaged over the three technology types and the three coal types.  The smallest plant that they analyse is 400MW, for this size plant the values given in \cite[Appendix A]{PPC1} are 11.6, 17.5, 8.9, 11.2, 16.9, 8.6, 11.0, 16.6 and 8.5 (\$$M$).  This gives an average value of \$$12.31M$ ($\pounds8.69 M$ inflation adjusted).  We need to adjust this to the capacity of the Drax Heron plant, a simple scaling would give $\pounds 8.69 M /400\times290=\pounds 6.30025$.  However the overheads for smaller plants are proportionally smaller, as can be seen from considering the 900MW plant use in Section~\ref{A:PPC}, this would give a scaled value of $\pounds 6.62167 M$.  The difference in capacity between the plants leading to these two figure is 500MW, and the closest of the figures comes from a plant which is 110MW larger than the plant we consider, therefore the overheads are given by $6.30- 110/500\times (6.622-6.300) = \pounds6.23 M$.

\subsection{Power Plant: Co-Fired}
The analysis of co-firing power plants (coal and bio-mass) is based on Drax,~\cite{PPCo1}, which has a capacity of 3960 MW and generates $24TWh$ of electricity each year.  Using the same selling price of electricity as both the coal-firing and biomass firing plants (the same as the anaerobic digester's unsubsidised price) of $\pounds 52.71/MWh$ gives revenue of $\pounds 1,265.04 M$.

As the plant burns both biomass and coal calculating the cost of materials requires knowledge of how much of each fuel is used each year.  From~\cite{PPCo1} we have that Drax consumes 9.1 M tonnes of coal a year and 1.725 M tonnes of biomass (a 660MW plant full burning biomass requires 2.3 M tonnes of biomass a year, whilst Drax co-fires 12.5\% biomass).   Using the same cost for coal as for the coal-firing power plant ($\pounds 19.05/tonne$) and the same cost for biomass as the biomass-burning power plant ($\pounds 50/tonne$) we arrive at a total material cost of $\pounds 259.56 M$.

We were unable to find any accurate data for overheads and production costs of co-firing plants.  Instead we use the values derived for the coal-firing power plant and multiply them by three, to scale from 900 MW to 3960 MW assuming a large economy of scale. 

\subsection{Composter}
There appears to be very little data on the economics of various types of composter; low grade, high grade and in-vessel.  Instead it seems that any economic analysis is on a case by case basis.  As such we do not treat all three of these `industry types' independently but rather simply as composters.  This means that the only difference in the composters in the model will be in their structural role in the metabolic interactions network.  We base our analysis of composters on~\cite{C1} which looks at a Canadian composter in 1993.  We do not believe that this should pose any issues to the accuracy of the figures derived as approximations due to the fact that composting is a well established technique, and whilst some efficiencies may have been made we do not believe that this would drastically effect the economics of a composter.   In what follows we use the standard notation C\$ to refer to Canadian dollars, also C\$1 in 1993 is now worth $\pounds 0.8774$ inflation adjusted.  The composter studied in \cite{C1} produces 1806 tonnes (3.254 ML) of compost a year,~\cite[Table 2]{C1}, which is sold at a price of C\$$0.053/L$ ($\pounds 0.0465$ inflation adjusted)~\cite[p.~2]{C1}.  This gives a revenue of $\pounds 151,318$.

For the material cost we use the same waste price as used in the analysis of an anaerobic digester, $\pounds41$. \cite[Table 2]{C1} gives that 2050 tonnes of waste is used each year, whilst \cite[p.~1]{C1} gives that bulking agents cost is between C\$12 and C\$13 per tonne of waste.  We use a value of C\$12.5 ($\pounds 10.97$ inflation adjusted.  This gives a total material price of -$\pounds 61,567$.  To calculate the overheads and production costs we use the annual operating expenses given in~\cite[Table 2]{C1}.  This gives a range of C\$45,423 - C\$79,920 including the cost of bulking materials.  Taking the average and removing the cost of the bulking materials gives an inflation adjusted value of $\pounds32,505$.  We assume that one third of this is overheads whilst the other two thirds are production costs.

\subsection{Waste Incinerator}
We base the analyse for waste incinerators on Newlincs~\cite{WI1} which burns 56,000 tonnes of waste a year to produce 3MW of electricity and 3MW of heat.  The incinerator operates for 8000 (full time equivalent) hours per annum, meaning that it produces 24GWh of electricity and of heat.  We use the same sale price for the electricity as the anaerobic digester, $\pounds 52.71/MWh$.  The same document that this figure came from also gives a value for the amount heat is sold for~\cite[Appendix A-5]{Extra3}; \euro{}$27.53/MWh$ ($\pounds 25.04$ inflation adjusted).  Thus the total revenue from the sale of electricity and heat is $\pounds1.866M$.  However there is another source of revenue that the incinerator receives, namely subsidies.  PFI credits~\cite{WI2} provide an additional revenue of $\pounds 40$ for every tonne of waste burnt, in total an additional $\pounds 2.24 M$.  We use this combined value for our analysis.  However PFI credits have recently (early 2013) been cancelled meaning that this subsidy is no longer being paid, this has caused legal action to be bought~\cite{WI3} by those affected.  Without this additional subsidy our economic analysis of waste incinerators shows them making a loss year on year and not being economically viable.

The material cost for waste incinerators is negative as they process waste from other industries.  We use the same figure for the cost of waste as we have used throughout our analysis, first mentioned in that of anaerobic digester, of $\pounds 41/tonne$.  Giving a total cost of materials of $-\pounds2.296 M$.   We use \cite{WI4} to calculate the overheads and production costs.  \cite[Table 10.11]{WI4} gives that the total costs are \euro{}$85.28/tonne$ ($\pounds 95.6/tonne$ inflation adjusted), and \cite[Table 10.45]{WI4} gives that the ratio between operational costs and overheads is 3.4:1.  We use operational costs as production costs meaning that production costs are $\pounds 4.14M$ and overheads are $\pounds 1.22M$.

\subsection{Landfill}
There are two different types of landfill sites in our model of the Humber region, those which deal with non-hazardous waste and those which deal with hazardous waste.  There appears to be very little data available for costs associated with hazardous landfill, and in particular the overheads and `production' costs.  As such we shall use a very rough estimate and say that they are twice as much as for non-hazardous landfill per tonne.  Both types of landfills are at the end point of the material chain meaning that they do not actually sell any product.   They only revenue generation is from the service they supply in taking waste from other industries - in our current classification this is the negative material cost.  As such the amount generated from sales for each of the landfill types is $\pounds 0$. 

For a non-hazardous landfill, we shall assume the processing of 80,000 tonnes of waste a year as this is the size or landfill analysed in~\cite{L1}.  Using the same cost for waste that we have used throughout ($\pounds 41/tonne$) gives a material cost of $\pounds-3.28 M$.  For hazardous waste we assume that a landfill processes 55,000 tonnes a year (this is the total hazardous waste processed in landfill in Yorkshire an the Humber)~\cite{L2}.  We use this figure as there are only a few landfills which are licensed to take hazardous landfill in the Humber region and they are not separate entities from those which process non-hazardous landfill.  We use a figure of $\pounds 85/tonnes$~\cite{L3} for the negative cost of the waste.  This means that the total material cost for a hazardous landfill is $\pounds 1.7M$. 

To calculate the overheads and production costs we use figures from~\cite[Table 8]{L1}.  For the overheads we use the construction costs (remaining phases (\$$5.70/tonne$) + contingency (\$$0.67/tonne$)), the site development costs (\$$0.19/tonne$), and the net interest on revenue bonds (\$$3.96/tonne$).  This gives overheads of \$$10.52/tonne$ ($\pounds 8.691/tonne$ inflation adjusted).  For the production costs we use the operating costs; \$$15.10/tonne$ ($\pounds 12.474/tonne$ inflation adjusted).  Multiplying these figures by the annual processing volume for the non-hazardous landfill and twice the annual processing volume for the hazardous landfill gives the values used in Table~\ref{T:Data}.

\subsection{Waste Aggregator}
In analysing waste aggregators we found that there was very little data available.  The data was particularly sparse for the smaller waste aggregators, this made it impossible to rigorously initialise the large regional and small local waste aggregators.  Instead we were forced to simply scale the values derived for a national waste aggregator.  The scaling we used can not be rigorously justified either but rather came as result of private correspondence~\cite{PC:FS} in which it was suggested that a national waste aggregator has a turnover of approximately $\pounds 1$ billion, whilst large regional waste aggregators have a turnover of between $\pounds 30 M$ and $\pounds 250 M$, and small local waste aggregators have a turnover of less than $\pounds 1 M$.  We therefore use a scaling for large regional waste aggregators of 10\% of national waste aggregators, and for small local waste aggregators of $0.1\%$ of national waste aggregators.

We base the values for a national waste aggregator on the annual accounts of Biffa~\cite{WA1}.  In 2011 Biffa had an annual revenue of $\pounds 775.1M$, a production cost of $\pounds 716.8 M$ (cost of sales), a distribution cost of $\pounds 22.6M$ and an administrative expense of $\pounds 26.9 M$, \cite[p.~16]{WA1}.  On \cite[p.~33]{WA1} is a list of the sources of revenue, this includes the money received for taking waste from various sources, which totals $\pounds 636 M$.  We use this value for the materials cost $-\pounds 636 M$ ($-\pounds 656.35 M$ inflation adjusted).  For the sales we consider the revenue less the money received for taking waste, that is $\pounds 139.1 M$ ($\pounds 143.55$ inflation adjusted).  The production cost is $\pounds 716.8 M$ ($\pounds 739.74 M$ inflation adjusted).  For the overheads we use the distribution costs and the administrative expenses, that is $\pounds 49.5 M$ ($\pounds 51.08 M$ inflation adjusted).
 
\subsection{Biodiesel Production: Waste}\label{S:BPW}
We base our analysis of biodiesel production on Brocklesby which has a turnover of between \$$20 M$~\cite{BPW1} and \$$30 M$~\cite{BPW2}.  We take the average of these two figures \$$25 M$ ($\pounds 16.25 M$).  
From private correspondence~\cite{INT1} with a representative from Brocklesby we use a volume of 20,000 tonnes as a
proxy for the final amount of biodiesel produced.  Converting this to litres gives $22.7 M L$ (density of biodiesel is 0.88~\cite{BPW3}).  Thus the capacity of Brockslesby is between 23 and 32~\cite{BPW4} million litres per annum.   We take the average value, to give an annual capacity of $27.5 ML$.  In~\cite[Table 6.9]{EXTRA1} the material cost is given as the range \$$0.16/L$ to \$$0.26/L$ (note that unlike the other waste streams we have previously considered for other industry types, due to the economic value of waste oils, this is actually a positive price).  Averaging this material cost gives \$$0.21/L$ ($\pounds 0.1495/L$ inflation adjusted), leading to a total material cost of $\pounds 4.11 M$.

\cite[Table 6.9]{EXTRA1} also gives an estimate for the production costs of \$$0.25/L$ ($\pounds 0.1755/L$ inflation adjusted), multiplying this by the annual capacity gives the figure used in Table~\ref{T:Data}.  For the overhead costs we use the overhead, packaging and storage costs as given in \cite[Table 4]{EXTRA2}, this gives the value for a plant with $9ML$ capacity of between \$$0.66M$ and \$$0.79M$.  Multiplying these values by three (to adjust for capacity) and averaging gives an overhead cost of \$$2.175M$ ($\pounds 4.82M$ inflation adjusted).

\subsection{Biodiesel Production: Virgin Feedstock}
We base the analysis of biodiesel production from virgin feedstock on our previous analysis for biodiesel production from waste, Section~\ref{S:BPW}.  For the revenue figure we use the top value found in Section~\ref{S:BPW} \$$30 M$ ($\pounds 19.5 M)$ instead of the average in order for our analysis of this industry type to show a profit.  We still use the same capacity as in the production from waste case, $27.5 ML$.  In order to find the material and production cost we again use \cite[Table 6.9]{EXTRA1} which gives a material cost of \$$0.53/L$ ($\pounds 0.377/L$ inflation adjusted), and a production cost of \$$0.3/L$ ($\pounds 0.2145/L$ inflation adjusted).  Multiplying these figures by the capacity gives the values shown in Table~\ref{T:Data}.  For the overheads we use the same value as that used in Section~\ref{S:BPW}.

\subsection{Chemical Production: Biological}
The analysis for chemical production is based on Croda.  All data is drawn from its 2012 annual statement~\cite{CP1}.  Croda had revenue of $\pounds 1,051.9 M$ and \emph{``consume[s] $\pounds 638.9 M$ of inventories"}~\cite[p.~86]{CP1} a year - we use this as the cost of materials.  To derive values for the overheads and production costs we use the figures for `cost of sales' and profits.  Croda's `cost of sales' are $\pounds 694.6 M$ and the profits are $\pounds 253.2 M$~\cite{CP1}.  For the overheads we use the revenue less the profit less the cost of sales, that is $\pounds 104.1 M$.  Finally for the production costs we use the cost of sales figure less the material cost, i.e.~$\pounds 55.7 M$.

\subsection{Bioethanol Production: Virgin Feedstock}
Our analysis of bioethanol production is based on Vireol which has a capacity to produce 200 million litres of bioethanol a year~\cite{BE1}.  Recent 12 month low and high spot prices for ethanol  were \$$2.16/gal$ and \$$2.76/gal$ respectively~\cite{BE2}, giving an average ethanol price of \$$2.46/gal$ ($\pounds0.422/L$) (one gallon - American - being 3.785 litres).  This gives an approximate revenue of $\pounds 84.5 M$.  To calculate the cost of materials we use~\cite[Table 5.6]{EXTRA1} which gives a feedstock cost of \$$0.5450/gal$ ($\pounds 0.1017/L$ inflation adjusted).  Multiplying this by the capacity gives a material cost of $\pounds20.33 M$.  

To find a value for the production costs we extrapolate from~\cite[Table 5.7]{EXTRA1} to arrive at a value of \$$0.37/gal$ ($\pounds0.069/L$ inflation adjusted).  Multiplying this by the capacity gives the desired value.  Also in \cite[Table 5.7]{EXTRA1} we find values for operating labour, SGA and maintenance of \$$0.04/gal$, \$$0.03/gal$ and \$$0.03/gal$ respectively.  We use these values to find a proxy for the overheads.  For biodiesel production the overheads were calculated as 60\% of operating labour, supervision and maintenance~\cite[Table 4]{EXTRA2}.  If we use this same ratio for bioethanol production, and use SGA as an approximation for supervision cost we arrive at a figure for overheads of \$$0.06/gal$ ($\pounds 0.011/L$ inflation adjusted).  Multiplying this value by the capacity of 200 million litres gives $\pounds 2.24 M$, the value shown in Table~\ref{T:Data}.

\subsection{Bioprocessor}
The values for a bioprocessor in Table~\ref{T:Data} are based on Cargill UK, Hull which uses 750 tonnes of rape seed a day to produce 420 tonnes of rape meal and 323 tonnes of rape oil~\cite{BP1}.  In the analysis that follows we shall assume that the plant operates for 300 full-time equivalent days a year.  We shall use $\pounds 280/tonne$ as the price for rape meal~\cite{BP2} and \$$1,150/tonne$ ($\pounds 748/tonne$) for rape oil (spot market, August 2013).  This means that the revenue generated is $\pounds 300\times ((420\times 280)+(323\times 748))=\pounds107.76M$.  Similarly a spot market price (August 2013) for rape seed of $\pounds 340/tonne$ gives the material cost shown in Table~\ref{T:Data}.

To calculate the overheads we use~\cite{BP3} which gives a value of \$$0.11/gal$ ($\pounds 0.0206/L$ inflation adjusted).  Working out the total overheads requires converting this value into a cost per tonne.  To do this we need the density of oil seed rape, which~\cite{BP4} gives as being between $0.711 Kg/L$ and $0.727 Kg/L$.  Taking the average density of $0.719 Kg/L$ we calculate total overheads of $\pounds 2.78 M$.  To calculate the production costs we use the processing costs of \$$0.5/gal$ ($\pounds£0.093/L$ inflation adjusted) given in~\cite[p.~26]{BP3}.  Again using the density of $0.719 Kg/L$ gives total production costs of $\pounds 12.57 M$.

\end{document}